\documentclass[11pt]{article}

\usepackage[dvips]{graphicx}
\usepackage{amssymb}
\usepackage{amsmath}
\usepackage{amscd}
\usepackage{mathrsfs}
\usepackage{latexsym}
\usepackage{makeidx}

\date{}
\author{Claudio Garola\footnote{Dipartimento di Fisica dell'Universit\`a del Salento and INFN, Sezione di Lecce; Via Arnesano, 73100, Lecce, Italy. E-mail: garola@le.infn.it.} \, and \, Sandro Sozzo\footnote{Dipartimento di Fisica dell'Universit\`a del Salento and INFN, Sezione di Lecce; Via Arnesano, 73100, Lecce, Italy. E-mail: sozzo@le.infn.it.}}
\title{\textbf{Reinterpreting Quantum Probabilities in a Realistic and Local Framework: The Modified BCHSH Inequalities}}

\begin{document}
\maketitle
\begin{abstract}
\noindent
Most physicists uphold that the tests of the Bell inequalities (BI) performed up to now confirm the predictions of standard quantum mechanics (SQM) and refute local realism. But some scholars criticize this conviction, defending local realism in various ways. We present here a new viewpoint based on an improved version of the \emph{extended semantic realism} (\emph{ESR}) \emph{model} that has been recently worked out by one of the authors. The ESR model embodies the mathematical formalism of SQM into a more general framework in which not only local realism but also objectivity of physical properties holds, and the probabilities of SQM are reinterpreted as \emph{conditional} instead of \emph{absolute}. Hence the ESR model provides some predictions that are formally identical to those of SQM but have a different physical interpretation, and further predictions that differ also formally from those of SQM. In particular, we show here that the BI introduced by Clauser, Horne, Shimony and Holt (\emph{BCHSH inequalities}) must be replaced by \emph{modified BCHSH inequalities}. These depend on \emph{detection probabilities} which may be such that the new inequalities are never violated by the \emph{conditional expectation values} predicted by the model. The condition that no violation occurs implies the existence of upper bounds on detection probabilities, which makes the ESR model falsifiable. These results admit an intuitive explanation in terms of \emph{unfair sampling} but basically differ from the seemingly similar results obtained by other approaches in which the \emph{efficiency problem} is discussed in order to vindicate some kind of local realism.

\vspace{.5cm}
\noindent
\textbf{Keywords.} Quantum mechanics; Bell inequalities; Local realism; Efficiency problem; Detection loophole.

\noindent
\textbf{PACS.} 03.65.\dag; 03.65.Ta; 03.65.Ud
\end{abstract}

\section{Introduction\label{intro}}
The term \emph{local realism} has been traditionally used to denote the joint assumptions of \emph{realism},

\emph{R: the values of all observables of a physical system in a given state are predetermined for any measurement context},

\noindent
and \emph{locality}, 

\emph{LOC: if measurements are made at places remote from one another on parts of a physical system which no longer interact, the specific features of one of the measurements do not influence the results obtained with the others}.

Assumptions R and LOC express local realism as it was originally intended in the literature \cite{b64,chsh69,epr35} (assumption LOC was usually stated referring to compound systems made up by two component parts only; we avoid introducing this restriction here since no part of our treatment requires it). They obviously hold in those \emph{local deterministic hidden variables theories} in which the values of the observables are determined by hidden parameters and do not depend on the measurement context. But assumption R introduces only a weak form of realism, since it does not require that the values of the observables be predetermined independently of the measurement context (there are indeed nonlocal, hence contextual, deterministic hidden variables theories, as Bohm's, in which R holds). Assumption LOC then limits the possible dependence on the measurement context by introducing the notion of part of a physical system and assuming that measurements on different parts of the same system must not influence each other if they are carried out far away. 

Assumptions R and LOC have been generalized in a stochastic sense in the literature dealing with the Einstein, Podolski and Rosen (EPR), or the EPR--Bohm, experiment. To be precise, R has been replaced by a stochastic form of realism, 

$R_S$: \emph{the probabilities of the values of all observables of a physical system in a given state are predetermined for any measurement context},

\noindent
and LOC has been replaced by a stochastic form of locality, 

$LOC_S$: \emph{if measurements are made at places remote from one another on parts of a physical system which no longer interact, the specific features of one of the measurements do not influence the probabilities of the results obtained with the others}.\footnote{We add that LOC and $LOC_{S}$ have been also replaced by a weaker assumption, 

\hspace*{.3cm}LOC$'$: \emph{no influence may be transmitted with a speed greater than that of light}, \\
which identifies locality with \emph{relativistic causality}, excluding only those influences that would conflict with special relativity (we remind that LOC$'$ has suggested several \emph{delayed choice} experiments, see, \emph{e.g.}, \cite{s04,s05}). \label{delayedchoice}} 

The term \emph{local realism} has then been also used with reference to \emph{objective local} \cite{ch74} or, equivalently, \emph{factorizable stochastic} \cite{f82,f89} models, which adopt $R_S$ and $LOC_S$ and add a \emph{factorizability assumption} F on probabilities (that, according to Fine, does not constitute a further locality condition, but requires that the correlations among the parts of a physical system in a given state be completely described by the hidden variables introduced by the model).

Assumptions $R_S$, $LOC_S$ and F seem jointly weaker than R and LOC. But, however intended, local realism leads to the Bell inequalities (BI) \cite{b64,chsh69,ch74,f82,f89}. It is then well known that the experimental tests \cite{g05} of the BI have produced a great number of data. These, according to most quantum physicists, show that the BI are violated and confirm the predictions of standard quantum mechanics (SQM),\footnote{We call \emph{standard quantum mechanics} here the Hilbert space formalism for quantum mechanics together with its orthodox (Copenhagen) interpretation. We stress the word \emph{standard} since we intend to propound a non--orthodox interpretation of the same formalism in this paper.} refuting local realism. 

Notwithstanding the large consent on the necessity of rejecting local realism, there are many scholars who criticize this conclusion. In particular, some authors object that the proofs of the BI rest not only on assumptions R and LOC, as usually stated, but on R and LOC together with a ``hidden Bell's postulate'' (HBP), which implies that a single probability measure exists serving for different experimental contexts. Therefore, the standard conclusion that R and LOC cannot hold simultaneously because SQM predicts violations of the BI must be substituted by the conclusion that R and LOC and HBP cannot hold simultaneously. Hence, one can reject HBP and avoid introducing hidden variables associated with remote observables, thus preserving LOC. In this way local realism would be recovered without introducing contradictions with SQM (a bibliography on this kind of approaches can be found in a recent review by Khrennikov \cite{k07}). Other authors accept instead that a conflict exists between SQM and local realism, but try to devise models that allow one to explain the obtained data without rejecting R and LOC (see, \emph{e.g.}, \cite{f89,dcg96,s00,sf02,gg99}). Should they be right, one could not decide whether SQM or local realism is correct on the basis of the experiments performed up to now.

The opposition of many scholars to rejecting local realism has deep reasons, which have been widely explored in the literature. Summing up, Santos \cite{s04} writes
\begin{quote}
``local realism is such a fundamental principle which should not be dismissed without extremely strong arguments''. 
\end{quote}
Indeed,
\begin{quote}
``as Einstein put it, without accepting the existence of an objective reality, independent of any observation, natural science would be impossible''. 
\end{quote}

Nevertheless, the attempts at recovering local realism mentioned above are not very popular among physicists. It is then relevant to observe that the local realism issue is strictly linked to another problem which is, in some sense, more fundamental, \emph{i.e.}, the problem of \emph{objectification} in SQM. The properties of a physical system, in fact, are usually maintained to be \emph{nonobjective} in SQM, which means that they cannot be considered as possessed or not possessed by the individual samples of the system independently of any measurement (one usually says that SQM is a \emph{contextual} theory). Nonobjectivity raises the question of how properties can be objectified by a measurement (see, \emph{e.g.}, \cite{blm91}; a broad bibliography on nonobjectivity and related topics can be found in this reference). This question has not yet received a satisfactory answer in the framework of SQM and its unsharp extensions \cite{bs96}. But it is apparent that, should an objective interpretation of the mathematical formalism of SQM be possible, not only the objectification problem would be avoided, but also R and LOC would hold in this interpretation. Moreover, this solution of the problem of recovering local realism would be basically different from the solutions mentioned above. Indeed, the authors who reject HBP retrieve local realism without refuting SQM, but necessarily introduce a kind of local contextuality (\emph{i.e.}, dependence on the local measurement context) which implies nonobjectivity, hence they leave open the problems following from this feature of SQM. The approaches that introduce models for explaining the experimental data in a framework in which R and LOC hold, instead, oppose local realism to SQM, hence do not provide any synthesis that allow one to reconcile these two fundamental components of our scientific thought.

The idea of recovering R and LOC as a byproduct of a reinterpretation of SQM that restores objectivity seems at first sight basically unsound because of the huge number of arguments which seem to show that nonobjectivity is an inherent feature of the quantum formalism.\footnote{There are at least two classes of no--go arguments of this kind. The first class is exemplified by the two--slit argument which is propounded in almost all manuals of SQM, and yet is deeply misleading in our opinion, since it mistakes physical states for physical properties, notwithstanding their different operational definitions (roughly speaking, we maintain that it is incorrect to claim that a sample of a physical system that is supposed to have a given physical property must be in the state in which this property is certain, see \emph{e.g.}, \cite{ga00}). Hence we do not take into account this class of arguments in this paper, even if they have been uphold and formalized in some valuable fundamental treatments of the quantum theory of measurement \cite{blm91}. The second class is exemplified by the Bell--Kochen--Specker (Bell--KS) theorem \cite{b66,ks67} which, according to most scholars, proves the contextuality of SQM (which holds true also at a distance because of the Bell theorem \cite{b64}), and we refer only to this class in the following. \label{twoslit}} Then, every attempt at implementing it must be based on a preliminary criticism of these arguments. Bearing in mind this remark, one of the authors, together with some coworkers, has proven in several previous papers (see, \emph{e.g.}, \cite{ga00,gs96a,gs96b,ga02b}) that the standard reasonings aiming to show that SQM conflicts with objectivity accept implicitly in SQM an epistemological assumption (called \emph{metatheoretical classical principle}, or, briefly, MCP), that does not fit in well with the operational philosophy of SQM itself. If MCP is replaced by a weaker \emph{metatheoretical generalized principle} (MGP), which is closer to the aforesaid philosophy, the proof of the conflict of SQM with objectivity cannot be given.\footnote{For the sake of completeness let us discuss this issue in more details. To this end, let us resume our epistemological position about physical laws. We consider the \emph{theoretical laws} of any physical theory as mathematical schemes from which \emph{empirical laws} can be deduced. Consistently with the operational and antimetaphysical attitude underlying quantum mechanics, we do not attribute truth values to the sentences stating the former. We instead assume that every sentence stating an empirical law has a truth value, which is \emph{true} in all those situations in which the law can be experimentally checked (\emph{epistemically accessible physical situations}), while it may be \emph{true} as well as \emph{false} in physical situations in which it cannot be checked because the theory itself prohibits any test. This assumption constitutes the basic content of the general principle that we have called MGP. If one then considers the canonical \emph{no--go theorems} mentioned in footnote \ref{twoslit}, one sees that they are proved \emph{ab absurdo}. To be precise, one assumes boundary, or initial, conditions which attribute noncompatible properties to the physical system that is considered. This implies hypothesizing physical situations that are not epistemically accessible. Nevertheless empirical quantum laws are applied in these situations \cite{ga02,gp04}, which subtends assuming in SQM that empirical laws are valid independently of the epistemic accessibility of the physical situation that is considered. This assumption, which is stronger than MGP, constitutes the basic content of the general principle that we have called MCP. If MCP is rejected and replaced by MGP, the above proofs cannot be carried out. \label{mgp}} Basing on this result and adopting MGP in place of MCP, the same author has supplied in the papers quoted above a new interpretation of the mathematical formalism of SQM according to which all properties associated with a physical system are objective.  The new interpretation obviously recovers local realism, adopts a purely semantic version of objectivity, hence of R\footnote{We note that R relates measurement apparatuses and (samples of) physical systems in a given state with values obtained whenever a measurement is performed, but it does not postulate the existence of some underlying ontological reality. Indeed, all entities mentioned in R (physical systems, observables, states) can be defined operationally. Hence, R can be interpreted in a purely semantic way, that is, as simply asserting that the truth value (\emph{true}/\emph{false}) of a sentence attributing a value of a physical observable to a given physical system is uniquely determined by the measurement context and by the state of the system. Similar remarks apply to objectivity, which then can also be interpreted in a purely semantic way.} (therefore it has been called \emph{Semantic Realism}, or \emph{SR}, \emph{interpretation}) and entails that the resulting theory is semantically incomplete (but all standard quantum laws hold formally unchanged in it, though their validity be limited by MGP). 

Notwithstanding its advantages, the SR interpretation may seem founded on a problematic epistemological analysis to many pragmatically oriented physicists. The same authors have therefore propounded an \emph{SR model} \cite{ga02,gp04} and a set--theoretical \emph{extended SR} (briefly, \emph{ESR}) \emph{model} \cite{ga02,gp04,ga03,ga05,gps06} that introduce a picture in which objectivity of properties, hence R and LOC, holds from the very beginning, MCP is falsified \cite{ga02} and there are intuitive reasons for assuming MGP. These models bring into every measurement a \emph{no--registration outcome} which is interpreted as providing information about the measured system and not as expressing the inefficiency of the measuring apparatus. In addition they incorporate the mathematical formalism of SQM together with its rules for calculating probabilities, but interpret the latter as providing \emph{conditional} instead of \emph{absolute} probabilities, which is consistent with MGP. To be precise, all probability values predicted by SQM are recovered and can be experimentally checked (in this sense one of us has sometimes written that the aforesaid models reconciliate local realism with quantum mechanics) but they are reinterpreted as referring only to the samples of the physical system which do not yield the no--registration outcome (conditional probabilities). These are not all samples, since the models assume that some samples may yield the no--registration outcome even if the measuring apparatus is perfectly efficient (idealized). Hence, whenever all samples are taken into account, one can get further predictions (regarding absolute probabilities) that are different from those of SQM and that, in principle, can also be experimentally checked.  

We intend to discuss in more details the ESR model in this paper and show that it constitutes a new theoretical scheme, which is still incomplete but already provides new results. More specifically, the content of the present paper can be resumed as follows.

\emph{Sect. \ref{modello}}. We remind the essentials of the ESR model and provide a new, more direct approach to it. We also introduce objectivity in this model by means of a new axiom which implies it (hence R and LOC).

\emph{Sect. \ref{consistenza}}. We discuss the possibility of generalizing the ESR model and the consistency of its axioms, and analyse the different kinds of statistical predictions that can be obtained from it.

\emph{Sect. \ref{misuremodello}}. We discuss some technical notions (expectation values, sequential measurements, correlation functions) in the ESR model that are needed to attain our results in the following sections.

\emph{Sect. \ref{disgeneralizzate}}. We consider the BI introduced by Clauser, Horne, Shimony and Holt (\emph{BCHSH inequalities}) from the viewpoint of the ESR model. In this model the quantum predictions obtained by using the rules of SQM hold whenever one takes into account only the samples of the physical system that are detected, hence the BCHSH inequalities are violated in this case. If, instead, one takes into account all samples that are produced, the ESR model predicts that \emph{modified BCHSH inequalities} hold. The new inequalities are never violated if suitable limits are imposed on the \emph{detection probabilities} that appear in them (which are determined by intrinsic features of the individual samples of the physical system, not by features of the measuring apparatuses or by the environment). Such limits can be empirically checked, at least in principle, which makes the ESR model falsifiable. 

\emph{Sect. \ref{unfairsampling}}. We comment  on the results obtained in Sect. \ref{disgeneralizzate}, and show that they follow because of the new interpretation of quantum probabilities introduced in the ESR model. We also provide an intuitive explanation of the violation of the BCHSH inequalities in terms of an unconventional kind of \emph{unfair sampling}.

\emph{Sect. \ref{confronto}}. We compare our approach with some approaches in the literature that regard the low detection efficiencies in the experiments (\emph{efficiency problem}) as a hindrance to interpreting experimental data as confirming SQM and refuting local realism. We stress that all these approaches accept the conflict of SQM with local realism without suggesting a more general perspective, as the ESR model does.

To close up, we point out that, notwithstanding the above results, we do not claim that the  ESR model provides a description of some kind of microscopic reality (though we do not reject this possibility). Rather, we maintain that it shows that the formal apparatus of SQM is compatible with objectivity, hence with local realism, contrary to a widespread belief. Thus a more manageable and paradox--free perspective can be constructed. The following statement by d'Espagnat \cite{desp98} illustrates our point properly.
\begin{quote}
``So, what I say is: concerning independent reality, perhaps one of these models - or some not yet discovered model - is right. We do not know and we shall never know. But the mere possibility that one is right obviously suffices to remove the difficulty''.
\end{quote}

\section{The ESR model\label{modello}}
As we have anticipated in Sect. \ref{intro}, the ESR model has been proposed by one of the authors few years ago \cite{ga02} and successively amended and improved in a number of papers \cite{gp04,ga03,ga05,gps06}. We present here a new version of it and introduce some comments that will be needed in the following sections.

The basic notions of the ESR model can be divided into three groups.

(i) Standard primitive and derived notions: \emph{physical system}, \emph{preparing device}, (\emph{pure}) \emph{state}, \emph{physical object}. In particular, a state of a physical system $\Omega$ is defined as a class of physically equivalent preparing devices \cite{bc81,l83}. A physical object is defined as an individual sample $x$ of $\Omega$, obtained by activating a preparing device $\pi$, and we say that ``$x$ is in the state $S$'' if $\pi \in S$.

(ii) New theoretical entities: \emph{microscopic properties}. We assume that every physical system $\Omega$ is characterized by a set $\mathcal E$ of microscopic properties. For every physical object $x$, the set $\mathcal E$ is partitioned in two classes, the class of properties that are possessed by $x$ and the class of properties that are not possessed by $x$, independently of any measurement procedure. We briefly say that microscopic properties are \emph{objective} (see Sect. \ref{intro}). We note explicitly that different physical objects in the same state $S$ may possess different microscopic properties (but assigning the state $S$ of $x$ imposes some limits on the subset of microscopic properties that can be possessed by $x$, see footnote \ref{certrue}). 

(iii) New observative entities: \emph{generalized observables}. We assume that every physical system $\Omega$ is associated with a set of generalized observables that also characterize it. Every generalized observable $A_0$ (here meant as a class of physically equivalent measuring apparatuses, without any reference to a mathematical representation) is obtained in the ESR model by considering an observable $A$ of SQM with spectrum $\Xi$ on the real line $\Re$ and adding a further outcome $a_0$ (\emph{no--registration outcome} of $A_0$) that does not belong to $\Xi$, so that the spectrum $\Xi_0$ of $A_0$ is given by $\Xi_0=\Xi \cup \{ a_0 \}$.

The introduction of generalized observables allows us to define the set ${\mathcal F}_{0}$ of all \emph{macroscopic properties} of $\Omega$,
\begin{equation}
{\mathcal F}_{0} \ =  \ \{ (A_0, \Delta), \  A_0 \ \textrm{generalized \ observable}, \ \Delta \ \textrm{Borel \ set \ on} \  \Re \},
\end{equation}
and the set ${\mathcal F} \subset {\mathcal F}_{0}$ of all macroscopic properties associated with observables of SQM,
\begin{equation} 
{\mathcal F} \ = \ \{ (A_0, \Delta), \  A_0 \ \textrm{generalized \ observable}, \ a_0 \notin \Delta \}.
\end{equation}
It is apparent that, for every generalized observable ${A}_{0}$, different Borel sets containing the same subset of $\Xi_{0}$ define physically equivalent properties. For the sake of simplicity we convene that, whenever we mention macroscopic properties in the following, we actually understand such classes of physically equivalent macroscopic properties. Furthermore, we agree to write simply \emph{observable} in place of \emph{generalized observable} whenever no misunderstanding is possible.

We establish a link between microscopic properties of $\mathcal E$ and macroscopic properties of $\mathcal F$ by means of the following assumption.\footnote{Microscopic properties do not appear in SQM. It is well known, however, that the attempt at interpreting macroscopic properties in $\mathcal F$ as properties of physical objects leads to the conclusion that such properties generally are \emph{potential} (hence nonobjective) and may become \emph{actual} (or objective) only whenever an ideal measurement is performed. In order to minimize the differences from the orthodox viewpoint, the general SR interpretation does not introduce microscopic properties and adopts an operational perspective \cite{gs96a,gs96b}, focusing on some inconsistencies with this perspective existing in the standard interpretation (see footnote \ref{mgp}) to avoid nonobjectivity and the paradoxes following from it. Also the SR model elaborated with the aim of showing the consistency of the SR interpretation \cite{ga02,gp04} does not introduce microscopic properties. On the contrary, the distinction between microscopic properties, which play the role of theoretical entities (see (ii)), and macroscopic properties, which play the role of observative entities (see (iii)), is crucial in the ESR model, since it allows one to supply an intuitive (set--theoretical) picture of the physical world. \label{microproperties}}

\emph{Ax. 1.} \emph{A bijective mapping $\varphi: {\mathcal E} \longrightarrow {\mathcal F} \subset {\mathcal F}_{0}$ exists}. 

Let us describe now an \emph{idealized} measurement by using the notions that we have introduced so far. 

Whenever a physical object $x$ is prepared in a state $S$ by means of a device $\pi$ and the observable ${A}_{0}$ is measured on it, the set of microscopic properties possessed by $x$ induces a probability (which is either 0 or 1 if the model is \emph{deterministic}) that the apparatus does not react, so that the outcome $a_{0}$ may be obtained. In this case, $x$ is not detected and we cannot get any explicit information about the microscopic properties possessed by it. If, on the contrary, the apparatus reacts, it yields the outcome $a \ne a_0$ if and only if the microscopic property $\varphi^{-1}((A_0, \{ a \}))$ is possessed by $x$. More generally, if the apparatus reacts, it informs us that the result lies in the Borel set $\Delta$, with $a_0 \notin \Delta$, if and only if $x$ possesses the microscopic property $\varphi^{-1}((A_0, \Delta))$ (for the sake of brevity, we say that $x$ possesses the macroscopic property $(A_0, \Delta)$ in this case). In addition, if $x$ possesses the macroscopic property $(A_0, \Delta)$, it also possesses all macroscopic properties of the form $(A_0, \Delta')$, with $\Delta \subseteq \Delta'$ and $a_0 \notin \Delta'$, and this occurs if and only if $x$ possesses all microscopic properties of the form $\varphi^{-1}((A_0, \Delta'))$. 

It follows from the above description that the result of an idealized measurement provides information on microscopic properties. Conversely, the microscopic properties determine the probability of a result (or the result itself if the model is deterministic), which therefore does not depend on features of the measuring apparatus (flaws, termal noise, etc.) nor is influenced by the environment. In this sense idealized measurements are ``perfectly efficient'', and must be considered as a limit of concrete measurements in which the specific features of apparatuses and environment must instead be taken into account.\footnote{For the sake of generality we do not introduce any assumption on the transformations of states induced by idealized measurements. When considering repeated measurements in the following (see Sect. \ref{misuremodello}) we define instead a subclass of idealized measurements for which such transformations are specified. \label{noprimaspecie}}

We must still place properly quantum probability in our picture. To this end, let us suppose that the device $\pi$ is activated repeatedly, so that a finite set $\mathscr S$ of physical objects in the state $S$ is prepared. Then, $\mathscr S$ can be partitioned into subsets ${\mathscr S}^{1}, {\mathscr S}^{2}, \ldots, {\mathscr S}^{n}$ such that in each subset all objects possess the same microscopic properties. We briefly say that the objects in ${\mathscr S}^{i}$ ($i=1,2, \ldots, n$) are in some \emph{microscopic state} $S^{i}$. This suggests us to associate every state $S$ with a (not necessarily finite) family of microscopic states $S^{1}, S^{2}, \ldots$ and characterize $S^{i}$ ($i=1,2, \ldots$) by the set of microscopic properties that are possessed by any physical object in ${S}^{i}$ (of course, also the microscopic states play then the role of theoretical entities in the ESR model). Let us now consider a physical object $x$ in $S^{i}$, and let us suppose that a measurement of a macroscopic property $F=(A_0, \Delta)$, with $a_0 \notin \Delta$, is performed on $x$ (which consists in testing whether the value of $A_0$ lies in the Borel set $\Delta$). Because of our description of the measurement process, whenever $x$ is detected, $x$ turns out to possess $F$ if and only if it possesses the microscopic property $f=\varphi^{-1}(F)$ (which occurs if and only if $f$ is one of the microscopic properties characterizing $S^{i}$). 
We are thus led to introduce the following probabilities.

${\mathscr P}_{S}^{i,d}(F)$: the probability that $x$ be detected when $F$ is measured on it.

${\mathscr P}_{S}^{i}(F)$: the conditional probability that $x$ turn out to possess $F$ when it is detected (which is 0 or 1 since $x$ either possesses $\varphi^{-1}(F)$ or not).

${\mathscr P}_{S}^{i,t}(F)$: the joint probability that $x$ be detected and turn out to possess $F$.

Hence, we get
\begin{equation} \label{formuladipartenza_i}
{\mathscr P}_{S}^{i,t}(F)={\mathscr P}_{S}^{i,d}(F) {\mathscr P}_{S}^{i}(F).
\end{equation}
Eq. (\ref{formuladipartenza_i}) is purely theoretical, since one can never know if a physical object is in the microstate $S^i$. Therefore, let us consider a physical object in the state $S$ and introduce a further conditional probability, as follows.

${\mathscr P}(S^{i}|S)$: the conditional probability that $x$, which is in the macroscopic state $S$, be in the microstate $S^i$.

The joint probability that $x$ be in the state $S^i$, be detected and turn out to possess $F$ is thus given by ${\mathscr P}(S^{i}|S) {\mathscr P}_{S}^{i,t}(F)$. Hence the joint probability ${\mathscr P}_{S}^{t}(F)$ that $x$ be detected and turn out to possess $F$ is given by
\begin{equation} \label{formuladipartenza_t}
{\mathscr P}_{S}^{t}(F)=\sum_{i}{\mathscr P}(S^{i}|S){\mathscr P}^{i,t}_{S}(F).
\end{equation}
Moreover, the probability ${\mathscr P}^{d}_{S}(F)$ that $x$ be detected when $F$ is measured is given by
\begin{equation} \label{formuladipartenza_d}
{\mathscr P}_{S}^{d}(F)=\sum_{i}{\mathscr P}(S^{i}|S){\mathscr P}^{i,d}_{S}(F).
\end{equation}
Let us define now
\begin{equation} \label{formuladipartenza_qm}
{\mathscr P}_{S}(F)=\frac{\sum_{i}{\mathscr P}(S^{i}|S){\mathscr P}^{i,t}_{S}(F)}{\sum_{i}{\mathscr P}(S^{i}|S){\mathscr P}^{i,d}_{S}(F)}.
\end{equation}
Then, we get
\begin{equation} \label{formuladipartenza}
{\mathscr P}_{S}^{t}(F)={\mathscr P}_{S}^{d}(F){\mathscr P}_{S}(F).
\end{equation} 
Eq. (\ref{formuladipartenza}) is the fundamental equation of the ESR model. Let us therefore discuss the two factors that appear in it.

Let us begin with the \emph{detection probability} ${\mathscr P}_{S}^{d}(F)$. We have seen that, since we are dealing here with idealized measurements, the occurrence of the outcome $a_0$ is attributed only to the set of microscopic properties possessed by $x$, which determines the probability ${\mathscr P}_{S}^{i,d}(F)$. Hence, ${\mathscr P}_{S}^{i,d}(F)$ neither depends on features of the measuring apparatus nor is influenced by the environment. Furthermore, the conditional probability ${\mathscr P}(S^{i}|S)$ depends only on $S$. Therefore, Eq. (\ref{formuladipartenza_d}) implies that ${\mathscr P}_{S}^{d}(F)$ depends only on the microscopic properties of the physical objects in $S$. But it must be noted that the ESR model does not provide a direct theoretical treatment of ${\mathscr P}_{S}^{d}(F)$, which thus remains indetermined (though some predictions on it can be obtained, as we show in the following). 

Let us come to ${\mathscr P}_{S}(F)$. By using Eqs. (\ref{formuladipartenza_i}) and (\ref{formuladipartenza_qm}) we get $0 \le {\mathscr P}_{S}(F) \le 1$. Moreover, the interpretations of ${\mathscr P}_{S}^{t}(F)$ and ${\mathscr P}_{S}^{d}(F)$ in Eq. (\ref{formuladipartenza}) show that ${\mathscr P}_{S}(F)$ can be interpreted as the conditional probability that a physical object $x$ turn out to possess the macroscopic property $F$ when it is detected. This interpretation of the term ${\mathscr P}_{S}(F)$ in Eq. (\ref{formuladipartenza}) provides a basis for the introduction of the main assumption of the ESR model. 

\emph{Ax. 2.} \emph{The probability ${\mathscr P}_{S}(F)$ can be evaluated by using the same rules that yield the probability $p_{S}(F)$ of $F$ in the state $S$ in SQM}.

Ax. 2 implies a new interpretation of the probabilities provided by standard quantum rules, which are now regarded as conditional rather than absolute, as we have anticipated in Sect. \ref{intro}. The old and the new interpretation of quantum probabilities coincide if ${\mathscr P}_{S}^{d}(F)=1$ for every state $S$ and property $F$. If there are states and properties such that ${\mathscr P}_{S}^{d}(F)<1$, instead, the difference between the two interpretations is conceptually relevant. Notwithstanding this, Ax. 2 preserves all standard quantum rules for evaluating probabilities, hence in particular the representation of states and macroscopic properties in $\mathcal F$ by means of trace class operators and (orthogonal) projection operators, respectively. The latter representation, however, does not apply to a macroscopic property $F_0=(A_0, \Delta)$ with $a_0 \in \Delta$, hence the (generalized) observable $A_0$ cannot be represented, as the observable $A$ of SQM from which it is obtained, by a self--adjoint operator. Moreover, no mathematical representation of microscopic properties and states is provided by SQM, hence by the ESR model.\footnote{Let us illustrate the difference between the old and the new interpretation of quantum probabilities with an example. In the literature on the foundations of quantum mechanics (see, \emph{e.g.}, \cite{bc81}) it is customary to associate every state $S$ of $\Omega$ with the set of all macroscopic properties that are \emph{certainly true} in $S$, \emph{i.e.}, have probability 1 for every physical object in the state $S$ according to SQM. In the ESR model the set ${\mathcal F}_{0S}$ of all properties that are certainly true in $S$ is different, and the model predicts that a physical object $x$ in the state $S$ possesses a property $F \in {\mathcal F}$ with probability 1 if and only if ${\mathscr P}_{S}^{d}(F)=1={\mathscr P}_{S}(F)$. It follows that the predictions of SQM do not coincide with those of the ESR model (of course similar reasonings apply when considering the set of all macroscopic properties that are \emph{certainly false} in $S$). In addition, we note that, if the physical object $x$ in the state $S$ is detected when the property $F \in \mathcal F$ such that ${\mathscr P}_{S}(F)=1$ is measured on it, $x$ necessarily possesses the microscopic property $f=\varphi^{-1}(F)$, which specifies the limits mentioned in (ii). \label{certrue}}

Let us observe now that the microscopic states can be seen as possible values of a \emph{hidden variable}. Hence, the ESR model provides a general scheme for a hidden variables theory, which exhibits some similarities with existing hidden variables theories or models but is different from all previous proposals because of its reinterpretation of quantum probabilities (we treat this topic in more details in Sect. \ref{confronto}). This viewpoint helps in discussing local realism and objectivity of the macroscopic properties in the ESR model (objectivity of microscopic properties is indeed assured by the basic assumptions of the model, see (ii)). To this end, let us consider Eq. (\ref{formuladipartenza_i}). Because of the definition of ${\mathscr P}_{S}^{i}(F)$, and since ${\mathscr P}_{S}^{i,d}(F)$ only depends on microscopic properties, this equation shows that the probability ${\mathscr P}_{S}^{i,t}(F)$ is completely determined by the values $S^{i}$ of the hidden variable (equivalently, by the set of all microscopic properties possessed by a physical object $x$ in $S^{i}$), hence it is independent of the measurement context. By considering a value $a \in \Xi_0$ of the observable $A_0$ and the property $F=(A_0, \{a \})$, it follows at once that assumptions $R_S$ and $LOC_S$ hold in the ESR model. In this sense local realism is preserved by this model. Moreover, in the special case of a deterministic ESR model, the result itself of any measurement is predetermined, so that all macroscopic properties are objective, hence R and LOC hold (see Sect. \ref{intro}). 

It remains to discuss objectivity of macroscopic properties whenever the ESR model is nondeterministic. We cannot deduce objectivity in this case, but we can introduce a further assumption which implies it, as follows.

\emph{Ax. 3.} \emph{For every microscopic state $S^{i}$, the probability ${\mathscr P}_{S}^{i,d}(F)$ admits an} epistemic (\emph{or} ignorance) \emph{interpretation} (see, \emph{e.g.}, \cite{bc81}) \emph{in terms of further unknown features of the physical objects in the state $S^{i}$}.

Because of Ax. 3, a parameter $\mu$ exists which determines, together with $S^{i}$, whether the physical object $x$ is detected whenever the property $F$ is measured on it, \emph{i.e.}, whether the outcome $a_0$ occurs or not (note that $\mu$ can be interpreted as denoting a subset of further microscopic properties possessed by $x$, selected in a new set of microscopic properties that do not correspond to macroscopic properties via $\varphi$). Since $S^{i}$ determines all macroscopic properties of $x$ whenever $x$ is detected, all macroscopic properties are determined by the pair $(\mu, S^{i})$, hence they are objective. In this way we obtain \emph{macroscopic objectivity} in the ESR model \cite{ga02,gp04,ga03,ga05}, which implies that R and LOC, not only $R_S$ and $LOC_{S}$, hold in it. 

\section{A consistency problem \label{consistenza}}
Our presentation in Sect. \ref{modello} shows that the ESR model can be considered as a new theoretical scheme, which embodies the mathematical formalism of SQM regarding the results of measurements and their probabilities and reinterprets the standard quantum probabilities, but is still incomplete. A minimal completion of it should introduce a mathematical representation of the (generalized) observables, provide rules for evaluating the detection probabilities and supply evolution laws. Of course such a completion should satisfy some reasonable requirements, which may pilot its construction. In particular, bearing in mind our epistemological analysis in Sect. \ref{intro}, one should adopt MGP (see footnote \ref{mgp}), hence generalize Ax. 2 assuming that the theoretical laws of the new theory must recover all empirical laws following from the formalism of SQM in all epistemically accessible physical situations (note that one can also adopt MGP independently of the ESR model and then deduce the interpretation of quantum probabilities as conditional \cite{gp04}). We stress that only a completion of this kind, not SQM, can be considered a universal theory from the viewpoint of the ESR model. In particular, one cannot deal with measurements in terms of interactions between a macroscopic apparatus and a microscopic physical object without resorting to a theory broader than SQM (a first step in this direction has been proposed in \cite{gp04}, where unitary evolution has been postulated when studying a measurement process in the framework of the SR model). This explains the failure of the attempts at providing an exhaustive theory of quantum measurements in SQM.\footnote{We have observed in Sect. \ref{intro} that unsharp quantum mechanics (UQM), though representing a substantial improvement of SQM, does not solve the objectification problem \cite{blm91,bs96}, as the ESR model aims to do. It is therefore interesting to point out a basic difference between the ESR model and UQM. To this end let us consider the projection operator $P_F$ and the density operator $\rho_S$ that represent the macroscopic property $F \in \mathcal F$ and the state $S$, respectively. Then, it follows from Eq. (\ref{formuladipartenza}) and from Ax. 2 that ${\mathscr P}_{S}^{t}(F)={\mathscr P}_{S}^{d}(F) Tr [\rho_S P_F]$, hence ${\mathscr P}_{S}^{t}(F)=Tr [\rho_S {\mathscr P}_{S}^{d}(F) P_F ]$. It is now evident that generalizing SQM by introducing effects instead of properties can never reproduce this formula, since the detection probability that appears in it depends on $S$ (see in particular, Eqs. (\ref{np})--(\ref{00})), while no such dependence appears in the expression of an effect.}

Leaving apart the above general problems and limiting ourselves to the ESR model presented in Sect. \ref{modello}, we may wonder whether its picture and axioms are consistent, since objectivity is rejected in SQM because of the theorems mentioned in footnote \ref{twoslit}. It is then apparent that the answer to this question depends on the choice of the detection probability ${\mathscr P}_{S}^{d}(F)$. Indeed, we have already seen that, if ${\mathscr P}_{S}^{d}(F)=1$ for any state $S$ and property $F$, the ESR model reproduces SQM, hence this choice of ${\mathscr P}_{S}^{d}(F)$ leads to inconsistencies. If instead ${\mathscr P}_{S}^{d}(F)<1$, one can show that MCP (see Sect. \ref{intro}) does not hold, hence the standard proofs of the foregoing theorems cannot be carried out \cite{ga02}. Nevertheless one cannot exclude that inconsistencies still occur. Therefore we should specify the general conditions that ${\mathscr P}_{S}^{d}(F)$ must fulfil in order to avoid contradiction. Unfortunately, we cannot supply these conditions at the present stage of our research, because of the aforesaid incompleteness of the ESR model. But we can pursue here a more modest aim, that is, we can consider standard cases in which the conflict between local realism and SQM is exhibited in the literature and show that in each of them the values of the detection probabilities can be assigned in such a way that no conflict occurs in the ESR model. These values can be experimentally checked, at least in principle,\footnote{In practice it may be hard to establish experimentally which predictions are correct because the intrinsic lack of efficiency of any measuring device hides the general term ${\mathscr P}_{S}^{d}(F)$. Indeed, the features of the measuring apparatus that may reduce the probability that $x$ be detected can be schematized by multiplying ${\mathscr P}_{S}^{d}(F)$ by a factor $k$, with $0 \le k \le 1$. Since the ESR model does not provide general theoretical predictions on ${\mathscr P}_{S}^{d}(F)$ it may be difficult to distinguish empirically $k$ from ${\mathscr P}_{S}^{d}(F)$, though this distinction is theoretically important (see Sect. \ref{disgeneralizzate}). \label{cappa}} which makes the ESR model falsifiable.
 
Of course we cannot take into account all cases that can actually be contrived. Therefore we limit ourselves in this paper to discuss the BCHSH inequalities (see Sect. \ref{disgeneralizzate}) that have played and play a fundamental role in the foundations of quantum mechanics, both from a theoretical and from an experimental viewpoint (hence we consider compound systems made up of two component subsystems only). Our treatment, however, provides a scheme which can be generalized for dealing with more complicate cases, as the proofs of nonlocality of SQM provided by Greenberger \emph{et al.} \cite{ghz90} or by Mermin \cite{m93} (where, in particular, compound systems made up of more than two component subsystems are considered), which we do not discuss here for the sake of brevity.

The above arguments are relevant if one considers the statistical predictions that can be obtained in the framework of the ESR model. Indeed, let a set of idealized measurements be performed on an ensemble of physical objects in a state $S$ (we remind that the ESR model deals with this kind of measurements only, hence the lack of efficiency of actual measuring apparatuses must be taken into account separately). It follows from the general features of the ESR model (in particular, Ax. 2) that its predictions can be partitioned in two classes.

(a) Predictions concerning the subensemble of physical objects that are detected by the measurements. They are obtained by using the quantum formalism (see Ax. 2), hence formally coincide with the predictions of SQM, but SQM would interpret them as referring to the ensemble of all objects that are produced. One expects that these predictions are matched by experimental data whenever idealized measurements are performed and only detected physical objects are considered. In this way the ESR model explains the outstanding success of the quantum formalism in providing accurate predictions about the natural world.

(b) Predictions concerning the ensemble of all objects in $S$. Here the detection probability ${\mathscr P}_{S}^{d}(F)$ plays an essential role, hence these predictions are mostly qualitative because of the lack of a general theory for ${\mathscr P}_{S}^{d}(F)$. Notwithstanding this, one can also obtain some quantitative predictions by requiring that the consistency conditions mentioned above be fulfilled, as we show in the next section.

\section{Expectation values and sequential measurements\label{misuremodello}}
The ESR model presented in Sect. \ref{modello} introduces a number of theoretical entities (microscopic properties and states) which have no operational definitions. But these entities do not appear in Eq. (\ref{formuladipartenza}), which can be postulated \emph{a priori} if one wants to reinterpret quantum probabilities without introducing underlying models. We therefore present some technical results in this section basing only on Eq. (\ref{formuladipartenza}), so that our arguments do not strictly depend on the ESR model (we show in Sect. \ref{unfairsampling}, however, that this model not only establishes a good background for attaining Eq. (\ref{formuladipartenza}), but also provides a set--theoretical intuitive justification of our achievements in Sect. \ref{disgeneralizzate}). 

Let us discuss firstly the expectation value of an observable $A_0$. For the sake of simplicity, let us assume that $A_0$ has discrete spectrum $\Xi_0= \{a_0 \} \cup \{ a_1, a_2, \ldots \}$ (the extension of our treatment to more general observables is straightforward). Measuring $A_0$ is equivalent to measuring the properties $F_0= (A_0, \{ a_0 \})$, $F_1= (A_0, \{ a_1 \})$, $F_2= (A_0, \{ a_2 \})$, \ldots simultaneously, so that, for every $F_n$ such that $n \in {\mathbb N}$, Eq. (\ref{formuladipartenza}) holds with $F_n$ in place of $F$, and we briefly write $a_n$ instead of $F_n$ in it. Furthermore, we consider the set of all observables for each of which the detection probability depends on the observable but not on its specific value, and introduce the reasonable physical assumption that this set is nonvoid (it could coincide with the set of all observables). We agree choosing the observables in this set from now on, and write ${\mathscr P}_{S}^{d}(A_0)$ instead of ${\mathscr P}_{S}^{d}(a_n)$, so that the probability of finding the outcome $a_n$ when measuring ${A}_0$ on $x$ in the state $S$ becomes
\begin{equation} \label{a_n}
{\mathscr P}_{S}^{t}(a_n)={\mathscr P}_{S}^{d}(A_0){\mathscr P}_{S}(a_n).
\end{equation}  
The probability ${\mathscr P}_{S}^{t}(a_0)$ of getting the outcome $a_0$ is instead given by
\begin{equation} \label{a_0}
{\mathscr P}_{S}^{t}(a_0)=1 - {\mathscr P}_{S}^{d}(A_0).
\end{equation}  

Eqs. (\ref{a_n}) and (\ref{a_0}) can be used to evaluate the expectation value ${\langle A_0 \rangle}_{S}$ of $A_0$ in the state $S$. We get
\begin{eqnarray} 
{\langle A_0 \rangle}_{S} = a_0 {\mathscr P}_{S}^{t}(a_0) + \sum_n a_n {\mathscr P}_{S}^{t}(a_n)= a_0 (1 - {\mathscr P}_{S}^{d}(A_0)) \nonumber \\
+ {\mathscr P}_{S}^{d}(A_0)\sum_n a_n {\mathscr P}_{S}(a_n)= a_0 (1 - {\mathscr P}_{S}^{d}(A_0))+ {\mathscr P}_{S}^{d}(A_0) {\langle A \rangle}_{S}. \label{expectation_value}
\end{eqnarray}
The term 
\begin{equation}
{\langle A \rangle}_{S}=\sum_{n}a_n {\mathscr P}_{S}(a_n)
\end{equation}
in Eq. (\ref{expectation_value}) deserves special attention. Indeed, because of Ax. 2 in Sect. \ref{modello}, it coincides with the standard quantum expectation value in the state $S$ of the observable $A$ of SQM from which $A_0$ is obtained, but its physical interpretation is different since it represents the mean value of $A_0$ whenever only detected physical objects are considered (see Sect. \ref{consistenza}, class of predictions (a)). Hence, we call it the \emph{conditional expectation value} of $A_0$ in the following.

We can then suppose, without loss of generality, that $a_0=0$ (hence, for every $n \in {\mathbb N}$, $a_n \ne 0)$. Indeed, whenever $a_0 \ne 0$, we can substitute $A_0$ with $\chi(A_0)$, where $\chi$ is a Borel function on $\Re$ which is bijective on $\Xi_0$ and such that $\chi(a_0)=0$, and $\chi(A_0)$ is defined as the observable obtained from $\chi(A)$ by adjoining the outcome $0$ and the detection probability ${\mathscr P}_{S}^{d}(\chi(A_0))={\mathscr P}_{S}^{d}(A_0)$. We get
\begin{equation}
{\langle A_0 \rangle}_{S} = {\mathscr P}_{S}^{d}(A_0) {\langle A \rangle}_{S}.
\end{equation}

Let now ${A}_0$ and ${B}_0$ be discrete observables with spectra $\{a_0 \} \cup \{ a_1, a_2, \ldots \}$ and $\{b_0 \} \cup \{ b_1, b_2, \ldots \}$, respectively, and let us agree to consider only idealized measurements of these observables that satisfy the following conditions.

(i) The measurement may change the state of the physical object $x$ on which it is performed but it does not destroy $x$, even if $x$ is not detected.

(ii) If the physical object $x$ is detected and a given outcome is obtained, the state of $x$ after the measurement can be predicted by using the projection postulate (more generally, the L\"{u}ders postulate) of SQM. 

We can then calculate the probabilities ${\mathscr P}_{S}^{t}(a_n,b_p)$, ${\mathscr P}_{S}^{t}(a_n,b_0)$, ${\mathscr P}_{S}^{t}(a_0,b_p)$, ${\mathscr P}_{S}^{t}(a_0,b_0)$ (with $n,p \in {\mathbb N}$) of obtaining the pairs of outcomes $(a_n,b_p)$, $(a_n,b_0)$, $(a_0,b_p)$, $(a_0,b_0)$, respectively, in a sequential measurement of ${A}_0$ and ${B}_0$ on a physical object $x$ in the state $S$. To this end, let us denote by $S_n$ the state of $x$ after a measurement of $A_0$ yielding outcome $a_n$ (which can be predicted by using the projection postulate of SQM because of (ii)), by $S_0$ the state of $x$ after a measurement of $A_0$ yielding outcome $a_0$ (that cannot be predicted by using the rules of SQM) and by ${\mathscr P}_{S}(a_n,b_p)$ the quantum probability of obtaining the pair $(a_n,b_p)$ when firstly measuring $A_0$ and then $B_0$. We get
\begin{eqnarray}
{\mathscr P}_{S}^{t}(a_n,b_p)={\mathscr P}_{S}^{t}(a_n){\mathscr P}_{S_n}^{t}(b_p)={\mathscr P}_{S}^{d}(A_0){\mathscr P}_{S}(a_n){\mathscr P}_{S_n}^{d}(B_0){\mathscr P}_{S_n}(b_p) \nonumber \\ 
={\mathscr P}_{S}^{d}(A_0){\mathscr P}_{S_n}^{d}(B_0){\mathscr P}_{S}(a_n,b_p), \label{np}
\end{eqnarray}  
\begin{equation} \label{n0}
{\mathscr P}_{S}^{t}(a_n,b_0)={\mathscr P}_{S}^{d}(A_0){\mathscr P}_{S}(a_n) (1-{\mathscr P}_{S_n}^{d}(B_0)),
\end{equation}
\begin{equation} \label{0p}
{\mathscr P}_{S}^{t}(a_0,b_p)=(1-{\mathscr P}_{S}^{d}(A_0)){\mathscr P}_{S_0}^{d}(B_0){\mathscr P}_{S_0}(b_p),
\end{equation}
\begin{equation} \label{00}
{\mathscr P}_{S}^{t}(a_0,b_0)=(1-{\mathscr P}_{S}^{d}(A_0))(1-{\mathscr P}_{S_0}^{d}(B_0)).
\end{equation}

The probabilities ${\mathscr P}_{S}^{d}(A_0)$, ${\mathscr P}_{S_n}^{d}(B_0)$, ${\mathscr P}_{S_0}^{d}(B_0)$, ${\mathscr P}_{S_0}(b_p)$ in Eqs. (\ref{np})--(\ref{00}) cannot be evaluated by using the rules of SQM. We can eliminate some of them by introducing a further selection of the idealized measurements that are considered by introducing a third condition, as follows.

(iii) The measurement is minimally perturbing in the sense that the state of a physical object $x$ is not changed by the measurement whenever $x$ is not detected.\footnote{Conditions (i)--(iii) define a subclass of idealized measurements (which we assume to be non--void). It is interesting to consider a measurement of an observable $A_0$ belonging to this class and wonder whether it can be classified as an \emph{ideal measurement of the first kind} in the sense established by SQM \cite{bc81,j68}. Let us therefore suppose that a first measurement is performed on a physical object in the state $S$ and then repeated on the physical object in the final state. If the first measurement yields outcome $a_n \ne a_0$, the second could yield $a_n$ as well as $a_0$; if the first measurement yields outcome $a_0$, the second could yield $a_n \ne a_0$, if the detection probability in the state $S$ is not 0. It follows that the measurement may yield a different result when repeated, hence it is not a measurement of the first kind. Furthermore, the outcome of the measurement determines the final state of the physical object, because of (i)--(iii), in such a way that the L\"{u}ders postulate can be extended to the outcome $a_0$ and is satisfied if and only if one associates the identical projection $I$ with the outcome $a_0$ (which is not orthogonal to the projection $P_n$ associated with an outcome $a_n \ne a_0$). In this sense an idealized measurement satisfying (i)--(iii) is ideal. More generally, the notion of ideality defined in some foundational approaches to SQM \cite{p76} applies to the idealized measurements considered here if the set ${\mathcal F}_{0S}$ of certainly true properties is introduced as in footnote \ref{certrue}. \label{idealispecie}}

Because of (iii), we get $S_0=S$, hence ${\mathscr P}_{S_0}^{d}(B_0)={\mathscr P}_{S}^{d}(B_0)$ and ${\mathscr P}_{S_0}(b_p)={\mathscr P}_{S}(b_p)$. Since the latter probability can be evaluated by using the rules of SQM, condition (iii) reduces the unknown probabilities to ${\mathscr P}_{S}^{d}(A_0)$, ${\mathscr P}_{S_n}^{d}(B_0)$ and ${\mathscr P}_{S}^{d}(B_0)$.

In addition, we observe that we are mainly interested in this paper to the special case of a compound physical system $\Omega$ made up by two far apart subsystems $\Omega_1$ and $\Omega_2$, with $A_0$ and $B_0$ observables of the component subsystems $\Omega_1$ and $\Omega_2$, respectively. Thus, we refer to this case from now on. Then, objectivity of properties implies that the change of the state of $\Omega$ induced by a measurement of $A_0$ on $\Omega_1$ must not affect the detection probability associated with the measurement of $B_0$ on $\Omega_2$.\footnote{Note that the objectivity of the properties belonging to $\mathcal F$ in the ESR model (see Sect. \ref{modello}) implies an epistemic interpretation of quantum probabilities. Hence the transition from a state $S$ to a state $S_n$ that occurs whenever a measurement of $A_0$ on a physical object $x$ yields outcome $a_n$, modifying our information about $x$, may change, for every $F \in \mathcal F$, the probability that $x$ possesses $F$, but does not necessarily change the (partially unknown) set of all properties possessed by $x$.} We therefore assume that ${\mathscr P}_{S_n}^{d}(B_0)={\mathscr P}_{S}^{d}(B_0)$. 

Because of the above assumptions, we get from Eqs. (\ref{np})--(\ref{00})     
\begin{equation} \label{assnp}
{\mathscr P}_{S}^{t}(a_n,b_p)={\mathscr P}_{S}^{d}(A_0){\mathscr P}_{S}^{d}(B_0){\mathscr P}_{S}(a_n,b_p),
\end{equation}  
\begin{equation}
{\mathscr P}_{S}^{t}(a_n,b_0)={\mathscr P}_{S}^{d}(A_0)(1-{\mathscr P}_{S}^{d}(B_0)) {\mathscr P}_{S}(a_n),
\end{equation}
\begin{equation}
{\mathscr P}_{S}^{t}(a_0,b_p)=(1-{\mathscr P}_{S}^{d}(A_0)){\mathscr P}_{S}^{d}(B_0){\mathscr P}_{S}(b_p),
\end{equation}
\begin{equation} \label{ass00}
{\mathscr P}_{S}^{t}(a_0,b_0)=(1-{\mathscr P}_{S}^{d}(A_0))(1-{\mathscr P}_{S}^{d}(B_0)),
\end{equation}
respectively. Eqs. (\ref{assnp})--(\ref{ass00}) contain only two probabilities that cannot be evaluated by using the rules of SQM, that is, ${\mathscr P}_{S}^{d}(A_0)$ and ${\mathscr P}_{S}^{d}(B_0)$. 

Following standard procedures and referring to the special case in which Eqs. (\ref{assnp})--(\ref{ass00}) apply, it is convenient for our aims to introduce also a \emph{generalized correlation function} $P(A_0,B_0)$, defined as follows.
\begin{eqnarray} 
P(A_0,B_0)=\sum_{n,p} a_n b_p {\mathscr P}_{S}^{t}(a_n,b_p)+\sum_{n} a_n b_0 {\mathscr P}_{S}^{t}(a_n,b_0) \nonumber \\
+ \sum_{p} a_0 b_p {\mathscr P}_{S}^{t}(a_0,b_p)+ a_0 b_0 {\mathscr P}_{S}^{t}(a_0,b_0). \label{corrfunc}  
\end{eqnarray}
By using Eqs. (\ref{assnp})--(\ref{ass00}), we get
\begin{eqnarray} 
P(A_0,B_0)=\sum_{n,p} a_n b_p {\mathscr P}_{S}^{d}(A_0) {\mathscr P}_{S}^{d}(B_0) {\mathscr P}_{S}(a_n,b_p) \nonumber \\
+ \sum_{n} a_n b_0 {\mathscr P}_{S}^{d}(A_0) (1-{\mathscr P}_{S}^{d}(B_0)) {\mathscr P}_{S}(a_n) \nonumber \\
+ \sum_{p} a_0 b_p (1-{\mathscr P}_{S}^{d}(A_0)){\mathscr P}_{S}^{d}(B_0) {\mathscr P}_{S}(b_p) \nonumber \\
+a_0 b_0 (1-{\mathscr P}_{S}^{d}(A_0))(1-{\mathscr P}_{S}^{d}(B_0)). \label{corrfuncgen}
\end{eqnarray}
Eq. (\ref{corrfuncgen}) can be simplified reasoning as above when dealing with the expectation value of $A_0$. Indeed one can choose, without loss of generality, $a_0=0=b_0$ (hence, for every $n,p \in {\mathbb N}$, $a_n \ne 0 \ne b_p$). Then, the generalized correlation function is given by
\begin{equation} \label{corrfunc_ass}
P(A_0,B_0)={\mathscr P}_{S}^{d}(A_0) {\mathscr P}_{S}^{d}(B_0) {\langle AB \rangle}_{S},
\end{equation}
where
\begin{equation} \label{corrfunc_QM}
{\langle AB \rangle}_{S}=\sum_{n,p} a_n b_p {\mathscr P}_{S}(a_n,b_p).
\end{equation}
Because of Ax. 2 in Sect. \ref{modello}, ${\langle AB \rangle}_{S}$ formally coincides with the standard quantum expectation value in the state $S$ of the product of the (compatible) observables $A$ and $B$ from which $A_0$ and $B_0$, respectively, are obtained. But its physical interpretation is different, as we have already observed referring to $\langle A \rangle_S$, and we call it the \emph{conditional expectation value of the product $A_0 B_0$}, consistently with the terminology introduced for $\langle A \rangle_S$.

As the standard correlation function in the literature, $P(A_0,B_0)$ may provide an index of the correlation among the outcomes that are different from $a_0$ and $b_0$. 

\section{The modified BCHSH inequalities\label{disgeneralizzate}}
It has been argued in a previous paper \cite{gp04} that the quantum violation of the BCHSH inequalities does not contradict local realism according to the SR and ESR models, since the quantum expectation values and the expectation values that appear in the BCHSH inequalities refer to different ensembles of physical objects in these models. Because of the specific aims of this paper, however, we think it appropriate to look into the subject in more details.

It is well known that the BCHSH inequalities are obtained by assuming R and LOC (see Sect. \ref{intro}). For the sake of brevity, we consider in the following only the original inequality provided by Clauser, Horne, Shimony and Holt \cite{chsh69}, that we write as follows,
\begin{equation} \label{chsh_69}
|P({\bf a}, {\bf b})-P( {\bf a}, {\bf b'})|+|P({\bf a'}, {\bf b})+P({\bf a'}, {\bf b'})| \le 2
\end{equation}
and briefly call \emph{standard BCHSH inequality} (of course, our reasonings apply to all BCHSH inequalities). The four terms on the left in inequality (\ref{chsh_69}) are \emph{correlation functions}, all of which are defined in the same way and differ only because of the choice of the parameters ${\bf a}$, ${\bf a'}$, ${\bf b}$, ${\bf b'}$. Let us therefore discuss only the first of them. This is given by
\begin{equation} \label{lhvt}
P({\bf a}, {\bf b})= \int_{\Lambda} d\lambda \rho(\lambda) A(\lambda, {\bf a}) B(\lambda, {\bf b}),
\end{equation}
where $\lambda$ is a \emph{hidden variable}, the value of which ranges over a domain $\Lambda$ when measurements on different samples of a physical system $\Omega$ in a given state $S$ are considered, $\rho(\lambda)$ is a probability distribution on $\Lambda$, ${\bf a}$ and ${\bf b}$ are fixed parameters, $A(\lambda, {\bf a})$ and $B(\lambda, {\bf b})$ are the values of two dichotomic observables $A({\bf a})$ and $B({\bf b})$, respectively, each of which can be $1$ or $-1$. Furthermore, $\Omega$ is assumed to be a compound physical system made up by two component subsystems $\Omega_1$ and $\Omega_2$, and $A({\bf a})$ and $B({\bf b})$ are observables of $\Omega_1$ and $\Omega_2$, respectively.

Let us resume now the orthodox viewpoint about inequality (\ref{chsh_69}). One considers all terms in the sum as expectation values of products of compatible observables. These values can be calculated in specific physical situations by using the rules of SQM. But if one puts them into inequality (\ref{chsh_69}), one easily sees that there are physical choices of the physical system, the parameters ${\bf a}$, ${\bf a'}$, ${\bf b}$, ${\bf b'}$ and the state $S$ which produce a violation of the inequality. One is thus led to think that the assumptions from which the inequality is deduced, \emph{i.e.}, R and LOC, are not consistent with SQM, which of course is a disconcerting conclusion that has puzzled physicists since 1964, when the first famous inequality was proven by Bell.

Let us come to the viewpoint introduced by the ESR model. Whenever only physical objects that are actually detected are taken into account, this model entails that the experimental data must fit in with the predictions that can be obtained by using the rules of SQM, hence there are physical situations in which the standard BCHSH inequality is violated. If, instead, all physical objects that are actually produced are taken into account, the ESR model entails that the experimental data must fit in with a new inequality that modifies the standard BCHSH inequality introducing in it the detection probabilities. To obtain this inequality, let us observe that in the ESR model the domain $\Lambda$ can be identified with the (discrete) subset of all microscopic states associated with the macroscopic state $S$ (see Sect. \ref{modello}) and $\rho(\lambda)$ with the conditional probability ${\mathscr P}(S^{i}|S)$ that a physical object is in the microstate $S^i$ whenever it is in the state $S$.\footnote{Equivalently, $\Lambda$ could be identified with a set of subsets of $\mathcal E$. Indeed, for every physical object $x$ in the state $S$, the parameter $\lambda \in \Lambda$ can be interpreted as denoting the set of all properties that are possessed by $x$. This identification has been adopted in a previous paper \cite{ga05}, where one of us has observed that microscopic properties (more properly, sets of microscopic properties) play the role of hidden parameters and must be distinguished from hidden variables in the standard sense since they do not fulfil the Kochen--Specker condition ``for the successful introduction of hidden variables'' \cite{ks67}. \label{microstates}} Furthermore, all macroscopic properties are objective, hence R and LOC hold, but a no--registration outcome must be adjoined to the spectrum of every observable (see again Sect. \ref{modello}). We can then follow the standard procedures used for getting inequality (\ref{chsh_69}), yet substituting the dichotomic observables $A({\bf a})$, $B({\bf b})$, $A({\bf a'})$, $B({\bf b'})$ by the trichotomic observables $A_0({\bf a})$, $B_0({\bf b})$, $A_0({\bf a'})$, $B_0({\bf b'})$, respectively, in each of which a no--registration outcome $0$ is adjoined to the outcomes $+1$ and $-1$. Thus, we write, in place of Eq. (\ref{lhvt}),
\begin{equation}
P(A_0({\bf a}),B_0({\bf b}))=\sum_{i} {\mathscr P}(S^{i}|S)A_0(S^i, {\bf a}) B_0(S^i, {\bf b}),  
\end{equation}
where $A_0(S^i, {\bf a})$ and $B_0(S^i, {\bf b})$ can be $1, -1$ and $0$. Since $|A_0(S^i, {\bf a})|\le 1$, we get
\begin{eqnarray}
|P(A_0({\bf a}),B_0({\bf b}))-P(A_0({\bf a}),B_0({\bf b'}))| \nonumber \\
 \le \sum_{i} {\mathscr P}(S^i|S) |B_0(S^i, {\bf b})-B_0(S^i, {\bf b'})|
\end{eqnarray}
and, similarly,
\begin{eqnarray}
|P(A_0({\bf a'}),B_0({\bf b}))+P(A_0({\bf a'}),B_0({\bf b'}))| \nonumber \\
\le \sum_{i} {\mathscr P}(S^i|S) |B_0(S^i, {\bf b})+B_0(S^i, {\bf b'})|.
\end{eqnarray}
Now, we have
\begin{equation}
|B_0(S^i, {\bf b})-B_0(S^i, {\bf b'})|+|B_0(S^i, {\bf b})+B_0(S^i, {\bf b'})| \le 2
\end{equation}
and
\begin{equation}
\sum_{i} {\mathscr P}(S^i|S)=1,
\end{equation}
hence we get
\begin{eqnarray} 
|P(A_0({\bf a}), B_0({\bf b}))-P(A_0({\bf a}), B_0({\bf b'}))| \nonumber \\
+ |P(A_0({\bf a'}), B_0({\bf b}))+ P(A_0({\bf a'}), B_0({\bf b'}))|\le 2. \label{chsh_gs}
\end{eqnarray}
By using Eq. (\ref{corrfunc_ass}) we finally obtain
\begin{eqnarray}
|{\mathscr P}_{S}^{d}(A_0({\bf a}))[{\mathscr P}_{S}^{d}(B_0({\bf b})){\langle A({\bf a})B({\bf b})\rangle}_{S}-{\mathscr P}_{S}^{d}(B_0({\bf b'})){\langle A({\bf a})B({\bf b'})\rangle}_{S} ]| \nonumber \\
+ |{\mathscr P}_{S}^{d}(A_0({\bf a'}))[{\mathscr P}_{S}^{d}(B_0({\bf b})){\langle A({\bf a'})B({\bf b})\rangle}_{S} \nonumber \\
+ {\mathscr P}_{S}^{d}(B_0({\bf b'})){\langle A({\bf a'})B({\bf b'})\rangle}_{S} ]|\le 2. \label{chsh_gsass}
\end{eqnarray}
The \emph{modified BCHSH inequality} (\ref{chsh_gsass}) replaces the standard BCHSH inequality (\ref{chsh_69}) in the ESR model. It contains explicitly four detection probabilities and four conditional expectation values. The latter can be calculated by using the rules of SQM because of the main postulate of the ESR model, and formally coincide with expectation values of SQM (see Sect. \ref{misuremodello}). If one puts them into inequality (\ref{chsh_gsass}), this inequality can be interpreted as a condition that must be fulfilled by the detection probabilities in the ESR model. Should one be able to perform measurements that are close to ideality, the detection probabilities could be determined experimentally\footnote{A major difficulty when performing an experiment for determining a detection probability is counting the physical objects that are actually produced, even if they are not detected by the measurement. Another difficulty is distinguishing the detection probability occurring because of intrinsic features of the physical object from the detection inefficiency occurring because of features of the physical apparatus (see footnote \ref{cappa}). Of course, we are only pointing out some theoretical possibilities here, and do not aim to suggest how measurements can actually be done. \label{experiments}} and then inserted into inequality (\ref{chsh_gsass}). Two possibilities occur.

(i) There exist states and observables such that the conditional expectation values violate inequality (\ref{chsh_gsass}). In this case one must reject the ESR model (hence R and LOC), or the additional assumptions introduced in Sect. \ref{misuremodello} in order to attain Eq. (\ref{corrfunc_ass}), or both.

(ii) For every choice of states and observables the conditional expectation values fit in with inequality (\ref{chsh_gsass}). In this case the ESR model is confirmed.

The above alternatives show that the ESR model is, in principle, falsifiable, as we have stated at the end of Sect. \ref{misuremodello}. Let us suppose that case (ii) occurs and that the ESR model is confirmed. Then, no conflict emerges between R and LOC, which hold in the model, and the reinterpreted quantum probabilities, which are embodied in it. 

The implications of inequality (\ref{chsh_gsass}) discussed above can be better understood by studying particular examples. Let us consider, for instance, a typical situation in the literature, in which $\Omega_1$ and $\Omega_2$ are two spin--$\frac{1}{2}$ quantum particles, $S$ is the singlet spin state represented by the unit vector
\begin{equation} \label{singlet}
|\eta\rangle=\frac{1}{\sqrt{2}} ( |+,-\rangle -|-,+\rangle),
\end{equation}
$A({\bf a})$ (or $A({\bf a'})$) is the observable ``spin of particle $\Omega_1$ along the direction ${\bf a}$ (or ${\bf a'}$)'' represented by the self--adjoint operator $\boldsymbol{\sigma}(1)\cdot {\bf a}$ (or $\boldsymbol{\sigma}(1) \cdot {\bf a'}$), and $B({\bf b})$ (or $B({\bf b'})$) is the observable ``spin of particle $\Omega_2$ along the direction ${\bf b}$ (or ${\bf b'}$)'' represented by the self--adjoint operator $\boldsymbol{\sigma}(2) \cdot {\bf b}$ (or $\boldsymbol{\sigma}(2) \cdot {\bf b'}$; for the sake of simplicity we have obviously omitted a factor $\frac{\hbar}{2}$ in the above representations). Then, it is well known that ${\langle A({\bf a})B({\bf b}) \rangle}_{S}= - {\bf a} \cdot {\bf b}$, and similarly ${\langle A({\bf a})B({\bf b'}) \rangle}_{S}= - {\bf a} \cdot {\bf b'}$, ${\langle A({\bf a'})B({\bf b}) \rangle}_{S}= - {\bf a'} \cdot {\bf b}$, ${\langle A({\bf a'})B({\bf b'}) \rangle}_{S}= - {\bf a'} \cdot {\bf b'}$ in SQM. In addition, the rotational invariance of the vector $|\eta\rangle$ and the choice of the observables suggest that the four detection probabilities in inequality (\ref{chsh_gsass}) must be identical in the case that we are considering. Let us therefore put ${\mathscr P}_{S}^{d}(A_0({\bf a}))={\mathscr P}_{S}^{d}(A_0({\bf a'}))={\mathscr P}_{S}^{d}(B_0({\bf b}))={\mathscr P}_{S}^{d}(B_0({\bf b'}))=p_{\eta}^{d}$. Then, we get from inequality (\ref{chsh_gsass})
\begin{equation} \label{bound}
(p_{\eta}^{d})^{2} \le \frac{2}{|{\bf a} \cdot {\bf b}- {\bf a} \cdot {\bf b'}|+|{\bf a'} \cdot {\bf b} + {\bf a'} \cdot {\bf b'}|} \, \, .
\end{equation}

Inequality (\ref{bound}) shows that the  ESR model predicts, under suitable assumptions, an upper bound for the probability that a spin--$\frac{1}{2}$ particle be detected when the spin along an arbitrary direction is measured on it and the compound system is in the singlet spin state. Since R and LOC hold in the model, this bound obviously cannot depend on the spin observables that are measured on the compound system, hence it coincides with the minimum value of the right member in inequality (\ref{bound}), which is $\frac{1}{\sqrt[4]{2}}\approx 0.841$ (the maximum value of the denominator in this inequality is indeed $2\sqrt{2}$). Thus, we get a prediction that, in principle, can be confirmed or falsified by means of experiments, even if it is difficult to imagine how this can be done because of the problems pointed out in footnote \ref{experiments}. It is interesting to observe that the above prediction also implies that the probability of getting the no--registration outcome when measuring $A_0({\bf a})$ (equivalently, $A_0({\bf a'})$, $B_0({\bf b})$, $B_0({\bf b'})$) has a lower bound in the special case considered here, which is $1-p_{\eta}^{d} \approx 0.159$.

Finally, we note that our procedures in this section establish a general paradigm for dealing with the BI in the ESR model, obtaining non--standard results and testable predictions.

\section{Conditional versus absolute probabilities: an intuitive explanation\label{unfairsampling}}
We have shown in Sect. \ref{disgeneralizzate} that the ESR model may lead to predictions that differ from those of SQM. We would like to discuss now in more details the features of the model that make this possible, also providing an intuitive picture of what may be going on at a microscopic level in the specific case discussed in Sect. \ref{disgeneralizzate}.

First of all, let us comment further on Eq. (\ref{formuladipartenza}) to avoid possible misunderstandings. To this end, let us consider a measurement of a physical property $F$ on a physical object $x$ in the state $S$, let us suppose that it is not perfectly efficient, and let us discuss it from the viewpoint of SQM. If we introduce an \emph{efficiency} $p_{S}^{d}(F)$ and denote the probability that $x$ has the property $F$ by $p_{S}(F)$, as in Ax. 2, the joint probability $p_{S}^{t}(F)$ that $x$ be detected and turn out to possess the property $F$ is given by
\begin{equation} \label{formulafine}
p_{S}^{t}(F)=p_{S}(F)p_{S}^{d}(F),
\end{equation}  
where $p_{S}(F)$ must be evaluated by using standard quantum rules, just as ${\mathscr P}_{S}(F)$ in Eq. (\ref{formuladipartenza}). The latter equation thus seems merely a restatement of Eq. (\ref{formulafine}) with different symbols. It is therefore important to stress that the physical interpretation of the two equations is neatly different. To better grasp this point, let us regard probabilities as large number limits of frequencies in ensembles of physical objects,\footnote{We adopt this na\"\i ve interpretation of physical probabilities here for the sake of simplicity. A more sophisticated treatment would associate quantum measurements with random variables, require that distribution functions approach experimental frequencies, etc. Our conclusions, however, would not be modified by the adoption of this more general and rigorous machinery.} and let us consider ensembles of physical objects in the state $S$. Then, the (conditional) probability ${\mathscr P}_{S}(F)$ (where $F=(A_0, \Delta)$, with $a_0 \notin \Delta$) in Eq. (\ref{formuladipartenza}) is the limit of the ratio between the number of objects in a given ensemble that are detected and possess the property $F$, and the number of objects that are detected. The (absolute) probability $p_{S}(F)$ in Eq. (\ref{formulafine}), instead, is the limit of the ratio between the number of objects in a given ensemble that possess the property $F$ and the number of objects in the ensemble. Hence, identifying the probability of $F$ in the state $S$ provided by quantum rules with ${\mathscr P}_{S}(F)$ instead of $p_{S}(F)$, as the  ESR model does, introduces a non--orthodox interpretation, as stated in Sect. \ref{modello}. This explains how we could reach some conclusions in Sect. \ref{disgeneralizzate} that do not agree with those of SQM. In fact, the expectation value of $A_0$ in Eq. (\ref{expectation_value}) is defined basing on our non--orthodox interpretation, while the expectation value of the observable $A$ from which $A_0$ is obtained does not depend in SQM on the efficiency of a concrete apparatus measuring $A$ (which usually varies with the outcome), so that one cannot use Eq. (\ref{formulafine}) in SQM to obtain an inequality analogous to inequality (\ref{chsh_gsass}).

The difference between the ESR model and the standard viewpoint can be appreciated even better by considering microscopic properties and states. Indeed, SQM introduces only macroscopic properties, that may be actualized if an ideal macroscopic measurement is performed (which implies, in particular, considering the efficiency as a ratio between the number of objects for which $F$ is concretely actualized and the number of objects for which $F$ would be actualized if the apparatus were ideal). The ESR model provides instead a set--theoretical picture of the microscopic world which makes its comparison with other theories easier. Therefore, let us briefly deal with this subject.

First of all, let us remind from Sect. \ref{modello} that, whenever an ensemble $\Sigma$ of physical objects is prepared in a state $S$, the microscopic properties possessed by each object depend on the microscopic state $S^{i}$ of the object (with some limits that have been specified in footnote \ref{certrue}) but all of them are objective, so that they do not depend on the measurement context. For every $f  \in \mathscr E$ one can then introduce a theoretical probability ${\mathscr P}_{S}(f)$ that a physical object $x$ in the state $S$ possesses $f$. Furthermore, let us consider the macroscopic property $F= \varphi (f)$ corresponding to $f$. The probability ${\mathscr P}_{S}(f)={\mathscr P}_{S}(\varphi^{-1}(F))$ does not coincide with the joint probability ${\mathscr P}_{S}^{t}(F)$ since, generally, there are physical objects that possess $f$ and yet are not detected, so that they do not possess $F$. Thus, ${\mathscr P}_{S}^{t}(F) \le {\mathscr P}_{S}(f)$. Instead, coming back to Eq. (\ref{formulafine}), we see that ${\mathscr P}_{S}(f)$ can be identified with the probability $p_{S}(F)$ introduced in this equation. Hence, from the viewpoint of the  ESR model the orthodox approach identifies the probability provided by quantum rules with ${\mathscr P}_{S}(f)$, while the model itself identifies it with ${\mathscr P}_{S}(F)$. 

The conceptual difference between the two perspectives is now clear. However, ${\mathscr P}_{S}(f)$ and ${\mathscr P}_{S}(F)$ need not be different. Indeed, two possibilities occur.

(i) The subensemble $\Sigma^{d}$ of all physical objects that are detected is a \emph{fair sample} of $\Sigma$, that is, the percentage of physical objects possessing $f$ in $\Sigma^{d}$ is identical to the percentage of physical objects possessing $f$ in $\Sigma$. Since all detected objects possessing $f$ turn out to possess $F=\varphi(f)$ when a measurement is done, ${\mathscr P}_{S}(f)$ and ${\mathscr P}_{S}(F)$ coincide.

(ii) $\Sigma^{d}$ is not a fair sample of $\Sigma$. In this case ${\mathscr P}_{S}(f)$ does not coincide with ${\mathscr P}_{S}(F)$. 

Our arguments can be generalized by introducing microscopic observables and their expectation values in the  ESR model, as follows.

Let $A_0$ be a discrete observable and let $\Xi_0 = \{ a_0 \} \cup \{ a_1, a_2, \ldots\} $ be the spectrum of $A_0$. Then, $A_0$ is characterized by the macroscopic properties $F_0= (A_0, \{a_0 \})$, $F_1= (A_0, \{a_1 \})$, $F_2= (A_0, \{a_2 \})$, \ldots (see Sect. \ref{misuremodello}). The property $F_0$ has no microscopic counterpart, while $F_1$, $F_2$, \ldots correspond to the microscopic properties $f_1=\varphi^{-1}(F_1)$, $f_2=\varphi^{-1}(F_2)$, \ldots, respectively. Then, we define the microscopic observable $\mathbb A$ corresponding to $A_0$ by means of the family $\{f_n \}_{n \in {\mathbb N}}$. The possible values of ${\mathbb A}$ are the outcomes $a_1, a_2, \ldots$ and its expectation value $\langle {\mathbb A} \rangle_{S}$ in the state $S$ is given by
\begin{equation}
\langle {\mathbb A} \rangle_{S}=\sum_{n} a_n {\mathscr P}_{S}(f_n),
\end{equation}
where ${\mathscr P}_{S}(f_n)$ is the theoretical probability of the microscopic property $f_n$. 

We are thus ready to discuss what is going on at a microscopic level. Indeed, by using the above definition we can consider the (dichotomic) microscopic observables ${\mathbb A}({\bf a})$, ${\mathbb A}({\bf a'})$, ${\mathbb B}({\bf b})$, ${\mathbb B}({\bf b'})$ corresponding to the (trichotomic) macroscopic observables $A_0({\bf a})$, $A_0({\bf a'})$, $B_0({\bf b})$, $B_0({\bf b'})$ introduced in Sect. \ref{disgeneralizzate}, respectively. Since all microscopic properties are objective, the usual procedures leading to inequality (\ref{chsh_69}) can be applied. Hence we get the standard BCHSH inequality, with $P({\bf a}, {\bf b})$, $P({\bf a}, {\bf b'})$, $P({\bf a'}, {\bf b})$, $P({\bf a'}, {\bf b'})$ reinterpreted  in terms of microscopic observables. 

Bearing in mind our results in Sect. \ref{disgeneralizzate}, we can now draw the interesting conclusion that different inequalities hold at different levels according to the ESR model. 

(a) The standard BCHSH inequalities hold at a microscopic level (which is purely theoretical and cannot be experimentally checked).

(b) The modified BCHSH inequalities hold at a macroscopic level whenever all physical objects that are actually produced are considered (which can be experimentally checked, at least in principle, see footnote \ref{experiments}).

(c) The quantum predictions deduced by using SQM rules hold at a macroscopic level whenever only detected physical objects are considered (which can be experimentally checked). In this case there are physical situations in which the standard BCHSH inequalities are violated, while quantum inequalities instead hold.\footnote{For instance, the inequality
\begin{displaymath}
|{\bf a} \cdot {\bf b}- {\bf a} \cdot {\bf b'}|+|{\bf a'} \cdot {\bf b} + {\bf a'} \cdot {\bf b'}| \le 2 \sqrt{2}
\end{displaymath}
which holds in the specific case studied at the end of Sect. \ref{disgeneralizzate} \cite{se88a,hs91}.}

To close up, let us suppose that $A_0$ is measured on each physical object in $\Sigma$. Then, several physical objects turn out to possess the property $F_0$ (hence the expectation value $\langle A_0 \rangle_{S}$ of $A_0$ is given by Eq. (\ref{expectation_value})). Therefore the objects for which the outcomes $a_1, a_2, \ldots$ are obtained belong to the subset $\Sigma^{d} \subseteq \Sigma$. Furthermore, the probabilities ${\mathscr P}_{S}(F_1)={\mathscr P}_{S}(a_1)$, ${\mathscr P}_{S}(F_2)={\mathscr P}_{S}(a_2)$, \ldots must be interpreted as the large number limits of the frequencies of $a_1, a_2, \ldots$, respectively, in $\Sigma^{d}$. Let us consider the conditional expectation value $\langle A \rangle_{S}=\sum_{n} a_n {\mathscr P}_{S}(F_n)$ introduced in Sect. \ref{misuremodello} and compare it with $\langle {\mathbb A} \rangle_{S}$. It is apparent that $\langle A \rangle_{S}$ and $\langle {\mathbb A} \rangle_{S}$ must coincide if case (i) above occurs, while they generally do not coincide if case (ii) occurs. Analogous remarks hold if we consider the conditional expectation value $\langle AB \rangle_S$ of the product of the observables $A_0$ and $B_0$ defined by Eq. (\ref{corrfunc_QM}). It follows that, if we substitute $P({\bf a}, {\bf b})$, $P({\bf a}, {\bf b'})$, $P({\bf a'}, {\bf b})$, $P({\bf a'}, {\bf b'})$ with conditional expectation values in the standard BCHSH inequality, this inequality must be fulfilled in case (i), while it can be violated in case (ii). Since the conditional expectation values coincide with standard quantum expectation values (see Sect. \ref{misuremodello}), there are physical situations in which the foregoing inequality is violated, hence we conclude that case (ii) occurs and $\Sigma^{d}$ is not a fair sample of $\Sigma$. We thus obtain a set--theoretical interpretation in terms of unfair sampling of the violation of the standard BCHSH inequality predicted by the ESR model.\footnote{This explanation of the violation of the standard BCHSH inequality was already provided in \cite{gp04}, where however only macroscopic properties were considered and the distinction between a macroscopic property $F$ and its microscopic counterpart $f= \varphi^{-1}(F)$ was not explicitly introduced. This made our argument somewhat ambiguous, and our present treatment also aims to amend this shortcoming. We add that unfair sampling obviously represent a necessary but not sufficient condition for the violation of inequality (\ref{chsh_69}), so that further quantitative conditions on it must be imposed if this violation has to occur. For the sake of brevity, we do not discuss this topic here.}

\section{The efficiency problem\label{confronto}}
We have underlined in Sect. \ref{intro} that our approach, hence the ESR model, is deeply different from the existing approaches that try to vindicate local realism by questioning the interpretation of the experimental results obtained up to now. But there are similarities between the ESR model and some earlier models in the literature. Moreover, the numerical bound in the example provided in Sect. \ref{disgeneralizzate} resembles similar bounds obtained by other authors. Hence one might be tempted to classify the ESR model as a new version of previous proposals. A more detailed comparison of it with other models can then be fruitful and prevent misunderstandings. 

To begin with, let us consider the following assumptions that can be found in some hidden variables models.

(i) The $0$ outcome is a possible result of an ideal measuring process.

(ii) The efficiencies of the detectors used in the experiments depend on the hidden variables.

Both assumptions (i) and (ii) appear, for instance, in \cite{f89}, \cite{s00} and \cite{sf02}. Fine explicitly states that ``no show'', coded as 0, is a possible result for an ideal measurement, and that ``underlying factors that act locally \ldots are presumed responsible \ldots also for the null result'' \cite{f89}. Also N. Gisin and B. Gisin introduce a ``no outcome at all'' as result of a measurement and use assumption (ii) in order to present a local hidden variables model which reproduces the correlations predicted by SQM (hence it explains the data obtained in the experiments) if the detector efficiency is not greater than 75\% \cite{gg99}. Only assumption (i) appears instead in other models. For instance, de Caro and Garuccio introduce ``no count events'', but explicitly reject assumption (ii), introducing a ``random nondetection'' in order to show that the results of the experiments can be compatible with local realism if the efficiency of the detectors is not greater than 0.811 \cite{dcg96}.

Coming to the ESR model, it is apparent that assumption (i) anticipates the introduction of the no--registration outcome as a possible result of idealized measurements in Sect. \ref{modello}. Assumption (ii) resembles the statement in Sect. \ref{modello} that the probability of detecting a physical object $x$ in a microstate $S^{i}$ depends on the microscopic properties of $x$, hence on $S^{i}$, which plays the role of value of the hidden variable in the ESR model (see Sect. \ref{modello}).

Let us discuss now why, notwithstanding the above similarities, the ESR model must be distinguished from the earlier models (of which the models quoted so far represent a limited sample). 

We preliminarily note that most scholars who argue against the interpretation of experimental data (ED) as a decisive refutation of local realism (see, \emph{e.g.}, \cite{s04,s05,se88a,hs91,se00,dcg94}) follow a common logical scheme. Indeed, they firstly observe that physicists do not actually test the BI but, rather, somewhat different inequalities (BI$^{*}$) that are obtained by adding additional assumptions (AA) (\emph{e.g.}, the ``renormalization of probabilities'', or the ``no--enhancement'', or the ``fair sampling'' assumption \cite{s04,s05}) to R and LOC. Then, they agree that the ED violate the BI$^{*}$ and oppose the standard conclusion that local realism must be rejected by means of a sequence of arguments that can be schematized, by using standard logical symbols, as follows.
\begin{eqnarray}
R \land LOC \land AA \Longrightarrow BI^{*}, \label{additionalassumptions1} \\
ED \Longrightarrow \lnot BI^{*}, \\
ED \Longrightarrow \lnot (R \land LOC \land AA), \\
ED \Longrightarrow \lnot (R \land LOC) \lor (\lnot AA). \label{additionalassumptions2}
\end{eqnarray} 
Implication (\ref{additionalassumptions2}) shows that the ED do not necessarily imply $\lnot$(R $\land$ LOC) (it holds, in particular, if (R $\land$ LOC) $\land$ ($\lnot$AA) is true), which challenges the common belief that the ED confirm SQM and refute local realism.

We stress now that the above conclusion does not imply that the ED are actually consistent with R and LOC in all performed experiments, nor explains why the ED match the predictions of SQM. Being aware of this, the authors of the models quoted at the beginning of this section complete the reasoning by denying the AA, introducing some specific hypotheses (SH) and showing that
\begin{equation}
(R \land LOC) \land (\lnot AA) \land SH \Longrightarrow ED
\end{equation}
(by the way, we underline the obvious implicit weakness of this procedure; in fact, the SH vary with the experiment, so that local realism is vindicated case by case introducing \emph{ad hoc} assumptions). 

Let us come to the arguments leading instead to the unified perspective expressed by the ESR model. These follow a different logical scheme. Indeed, let us firstly note that the usual reasonings aiming to prove that SQM conflicts with local realism can be schematized as follows (see also \cite{gp04}).
\begin{eqnarray}\label{bellinequalities}
(R \land LOC) \Longrightarrow BI, \\ 
SQM \Longrightarrow \lnot BI, \\
SQM \Longrightarrow \lnot(R \land LOC), \label{nonlocality}
\end{eqnarray}
(we add that implication (\ref{nonlocality}) is equivalent to SQM $\Longrightarrow$ ($\lnot$R) $\lor$ ($\lnot$LOC); thus one usually completes the reasoning by observing that SQM $\land$ ($\lnot$R) $\Longrightarrow$ $\lnot$LOC and concluding that SQM $\Longrightarrow$ $\lnot$LOC, see, \emph{e.g.}, \cite{r87} and \cite{g95}). Then, let us remind that SQM interprets quantum probabilities as absolute (see Secs. \ref{intro} and \ref{modello}), that is, as referring to the set of all physical objects that are prepared. Let us denote this interpretation by AP, so that
\begin{equation}
SQM \Longleftrightarrow (QM' \land AP)
\end{equation}
(where QM$'$ simply denotes SQM without AP). Hence the sequence (\ref{bellinequalities})--(\ref{nonlocality}) can be restated by substituting SQM with QM$'$ $\land$ AP in it. The last implication that is thus obtained is equivalent to
\begin{equation}
(R \land LOC) \Longrightarrow (\lnot QM' \lor \lnot AP)
\end{equation}
which shows that R $\land$ LOC needs not contradict QM$'$, since it could simply contradict AP. Thus, the ESR model can be contrived, which substitutes AP with a new interpretation of quantum probabilities as conditional instead of absolute, that is, as referring to the subset of physical objects that are detected.

The difference between the models mentioned at the beginning of this section and the ESR model is now evident. As we have anticipated in Sect. \ref{intro}, the former accept that SQM conflicts with local realism, hence with the BI, but stress that the ED actually violate the BI$^{*}$, not the BI, which is not sufficient for claiming that SQM holds and local realism is disproved. Then they show that the ED can actually be explained in frameworks in which local realism holds. The ESR model entails instead that the ED, whenever one takes into account only those physical objects that are detected, must match the same predictions made by SQM with reference to the set of all physical objects that are prepared, hence must fulfil quantum predictions and violate the BI (in particular, the standard BCHSH inequality) in specific physical situations. But this violation is compatible with local realism and no contradiction occurs.\footnote{Santos argues in some recent papers \cite{s04,s05} for the possibility of modifying SQM, without destroying its formalism and its impressive agreement with experiments, in such a way that local realism and SQM become compatible. In our opinion the SR interpretation and the ESR model just realize this program, as we have illustrated in the present paper and one of us has expounded in a huge number of articles going back to 1994.}

The above result seems relevant to us. But it must not be forgotten that it has a price. Indeed, we have reminded in Sect. \ref{intro} that MGP instead of MCP holds in the ESR model \cite{ga02}. This implies that one cannot consider empirical quantum laws as valid in physical situations that are \emph{in principle} not accessible to empirical tests (see footnote \ref{mgp}; this restriction was not stated explicitly when presenting the ESR model in Sect. \ref{modello} since it is implicit in the reinterpretation of quantum probabilities provided by the model).


\end{document}